\documentclass[conference, letterpaper, 10pt]{IEEEtran}
\usepackage[utf8]{inputenc}

\usepackage[backend=biber,
    bibstyle=ieee,
    sortcites=true,
    maxnames=2,
    maxbibnames=12,
    minnames=1
]{biblatex}

\usepackage{xspace}
\newcommand*{\eg}{e.g.\@\xspace}
\newcommand*{\ie}{i.e.\@\xspace}

\usepackage{hyperref}

\usepackage{amsmath,amssymb,amsfonts}
\usepackage{algorithmic}
\usepackage{graphicx}
\usepackage{textcomp}
\usepackage{xcolor}
\usepackage{csquotes}
\usepackage{booktabs}
\usepackage{tabularx}
\definecolor{sron0}{HTML}{332288}
\definecolor{sron1}{HTML}{88CCEE}
\definecolor{sron2}{HTML}{117733}
\definecolor{sron3}{HTML}{DDCC77}
\definecolor{sron4}{HTML}{CC6677}
\definecolor{sron5}{HTML}{AA4499}
\usepackage{subfig}

\usepackage{incgraph,tikz}
\usepackage{pgfplots}
\usetikzlibrary{
        pgfplots.dateplot,
        pgfplots.groupplots,
        external,
    }
\pgfplotstableread[col sep=comma]{./countries.csv}\datatable
\pgfplotstableread[col sep=comma]{./countries_block.csv}\datatabl
\pgfplotstableread[col sep=comma]{./countries_ip.csv}\datatab
\pgfplotstableread[col sep=comma]{./small_1MB_country.csv}\datatabelle

\usetikzlibrary{calc,fit}
\usetikzlibrary{shapes,positioning,chains}
\usetikzlibrary{decorations.pathreplacing}
\usetikzlibrary{decorations.text}
\usetikzlibrary{patterns}
\usetikzlibrary{backgrounds}

\begin{document}

\title{Map-Z: Exposing the Zcash Network\\in Times of Transition}

\author{
\IEEEauthorblockN{Erik Daniel}
\IEEEauthorblockA{\textit{Distributed Security Infrastructures} \\
\textit{Technical University of Berlin}\\
erik.daniel@tu-berlin.de}
\and
\IEEEauthorblockN{Elias Rohrer}
\IEEEauthorblockA{\textit{Distributed Security Infrastructures} \\
\textit{Technical University of Berlin}\\
elias.rohrer@tu-berlin.de}
\and
\IEEEauthorblockN{Florian Tschorsch}
\IEEEauthorblockA{\textit{Distributed Security Infrastructures} \\
\textit{Technical University of Berlin}\\
florian.tschorsch@tu-berlin.de}
}
\maketitle

\begin{abstract}
Zcash is a privacy-preserving cryptocurrency that provides anonymous monetary transactions.
While Zcash's anonymity is part of a rigorous scientific discussion,
information on the underlying peer-to-peer network are missing.
In this paper, we provide the first long-term measurement study of the Zcash network
to capture key metrics such as the network size and node distribution
as well as deeper insights on the centralization of the network.
Furthermore, we present an inference method based on a timing analysis of block arrivals
that we use to determine interconnections of nodes.
We evaluate and verify our method through simulations and real-world experiments,
yielding a precision of 50\,\% with a recall of 82\,\% in the real-world
scenario.
By adjusting the parameters, the topology inference model is adaptable to the
conditions found in other cryptocurrencies and therefore also contributes to
the broader discussion of topology hiding in general.
\end{abstract}
 
\vspace{-0.5em}
\section{Introduction}\label{sec:intro}
Since Bitcoin's~\cite{nakamoto2008bitcoin} introduction in 2008, a large
number of so-called altcoins emerged.
Altcoins are inspired by the general principle of a distributed
ledger, but aim to improve the system design in one way or another.
One of these altcoins is the Zcash project, which implements the Zerocash payment
scheme~\cite{ben2014zerocash}. Zcash launched in October 2016
and aims to enable anonymous, yet publicly verifiable, transactions
by employing zero-knowledge proofs~\cite{bensasson2014succinct}.

To date, most research on Zcash concentrates on anonymity and security aspects
of the ledger~\cite{quesnelle2017linkability,kappos2018empirical}.
Another central aspect of the system, however, has not yet been
considered in prior investigations: the Zcash peer-to-peer network.
This is particularly notable, since it has been shown before that the
properties of the networking layer may have
serious consequences on the security and privacy of blockchain-based systems~\cite{croman2016scalingblockchains,gervais2016security}.
Moreover, revealing the network topology may render even the best privacy-preserving cryptographic primitives
ineffective, if it is possible to link transactions and blocks to specific
peers~\cite{biryukov2014deanonymisation,fanti2017deanon}.
On a general note, monitoring the network status and health can indicate arising issues,
such as a high degree of mining power centralization.
This sort of centralization can be considered critical as it puts
the majority of computational resources in the hands of a single entity, which
could empower a malicious actor to rig the consensus in her favor.
Centralization in general undermines the goal of distributed trust
since the payment history could be altered at any time~\cite{nakamoto2008bitcoin}.

In this paper, we provide the first longitudinal measurement study of
the Zcash peer-to-peer network and present a model for topology inference.
In our measurement study, we expose information about Zcash nodes and their connectivity.
To this end, we observed the Zcash peer-to-peer network from a vantage point
over a period of four month (July to October 2018), covering multiple major client and protocol updates.\footnote{The measurement data and simulation code base is publically
available at \url{https://gitlab.tubit.tu-berlin.de/erik\_1105/map-z\_data}}
The vantage point is a deployed reference client, modified for data collection.
The empirical data allows us to conduct a global analysis of the Zcash network,
capturing key metrics such as network size, block propagation
behavior and timing, mining power distribution, and node churn.
We also use the data to reveal information about the nodes themselves,
including client version and global node distribution.

Based on the measured data of the Zcash network and
previous ideas for topology inference in the Bitcoin network~\cite{neudecker2016timing},
we developed a passive topology inference method which exploits the block relay mechanism.
While our method allows us to infer connections of mining nodes and their direct neighbors only,
it otherwise has a number of advantages:
It is independent of countermeasures for active timing analyses based on
transactions, such as trickling or diffusion spreading~\cite{fanti2017deanon}.
Additionally, the usage of blocks as a basis for inference produces no costs in terms of transaction fees.
We verify our model in a simulation using the Boost Graph Library~\cite{bgl}
and present various ways to determine the parameters of our model.
Moreover, we test the approach in a series of real-world experiments with two
Zcash network nodes.
In this scenario, the model could achieve a precision of 50\,\% and
a recall of 82\,\%, proving its applicability in a real-world setting.
In general, we were able to reproduce and improve the results of~\cite{neudecker2016timing}.
As the topology inference model is adaptable to the protocol specifications of other cryptocurrencies, it contributes to the discussion of topology hiding in general.

The rest of the paper is structured as follows. Section~\ref{sec:relwork} shows similar research concerning characterization and topology inference of cryptocurrencies. Section~\ref{sec:mapping} presents our research of the Zcash peer-to-peer network.
In Section~\ref{sec:inference}, we introduce our inference model which is evaluated in Section~\ref{sec:evaluation}. Section~\ref{sec:conclusion} concludes the paper.

\section{Related Work}\label{sec:relwork}
So far, most research in the field of cryptocurrencies concentrates on
Bitcoin and Ethereum: prior contributions~\cite{decker2013propagation,
pappalardo2018blockchain,kim2018measuring} explored both networks and
collected statistical data
by deploying measurement nodes. Among others, they observed that the block
propagation delay has a significant impact on the security and fairness of the
consensus layer. This has also been studied analytically and with the help of
simulations~\cite{croman2016scalingblockchains, gervais2016security}.
In~\cite{gencer2018decentralization}, the degree of centralization of Bitcoin and Ethereum is investigated.
The authors found both networks to be highly centralized in terms of (geographical) mining power
distribution. As also shown in~\cite{kaiser2018looming}, the mining power distribution is
dominated by Chinese mining pools, which bears the risk of intervention by a
state actor. Moreover, the authors argue that the interference of the
so-called \enquote{Great
Firewall of China} leads to degraded block propagation performance in the
Bitcoin network.

Beyond Bitcoin and Ethereum, Zcash exhibits similar properties to Bitcoin
but provides additional privacy guarantees by implementing so-called shielded transactions,
which use zero-knowledge proofs~\cite{bensasson2014succinct} to yield anonymous transactions.
Most prior work on Zcash focuses on the anonymity aspects.
In~\cite{quesnelle2017linkability}, the authors
studied the effectiveness of the anonymity features in Zcash by analyzing the
shielded function usage patterns. Similarly, the authors of~\cite{kappos2018empirical}
analyzed the data stored in the Zcash blockchain
to evaluate its anonymity. However, to the best of our knowledge,
no previous work studied the network behavior of Zcash.
Our work therefore provides the first measurement study
concentrating on Zcash's peer-to-peer networking layer.

Network tomography in general comprises methods to infer internal network characteristics,
including topology inference.
The methods basically differ in the employed type of probing technique,
which boils down to either unicast or multicast measurements.
While unicast measurements, such as sandwich
probes~\cite{coates2002maximum} or RTT measurements~\cite{shirai2007fast}, are
hard to realize in Zcash, as the observer cannot determine the spreading path,
multicast measurements are easier to reproduce.
Most approaches using multicast measurements, \eg,~\cite{duffield2002multicast},
use end-to-end loss instead of delay, though.
Since loss rates can only be measured between two peers,
these approaches are not applicable in Zcash as well.

There exists research that concentrates on topology inference of
cryptocurrencies: \citeauthor{miller2015discovering}~\cite{miller2015discovering}
use the timestamps
of announced IP addresses to infer links between different nodes and also try
to find mining pools and influential nodes.
Another approach by \citeauthor{grundmann2018exploiting} is primarily used to expose only a few targeted nodes.
The approach uses a method to detect the topology of Bitcoin by exploiting its transaction
propagation mechanics~\cite{grundmann2018exploiting}.

In~\cite{neudecker2016timing}, \citeauthor{neudecker2016timing} present a
method to infer the topology of a peer-to-peer flooding network like Bitcoin
with the help of an analytical model. The authors
list in~\cite{neudecker2019network} topology hiding as a security
requirement for blockchain networks and present an adversary model
which tries to infer the topology based on the peer discovery mechanism.
Furthermore, they show the importance of the transaction relay delay as a countermeasure
against topology inference. The ideas of both papers serve as the basis for
the conducted topology inference.
In this paper, our inference model however extends the approach by employing
a passive, cost-efficient monitoring, and more accurate approach, promising to capture the whole network.
 \section{Mapping the Zcash Network}\label{sec:mapping}
In this section, we present results of our measurement study of the Zcash network.
The goal of this investigation is to illuminate network characteristics on a global scale.
To this end, the empirical data covers two main categories:
data about individual nodes and about information propagation between nodes.
In particular, we show the worldwide interest in Zcash
and present insights on
network size and stability,
geographic node distribution,
software deployment and lifecycles,
origin of blocks,
block propagation time,
as well as mining centralization.

\begin{figure*}
\begin{tikzpicture}
\footnotesize
\begin{axis}[
  xlabel=Time,
  ylabel=Number of Connections,
  date coordinates in=x,
  date ZERO=2018-07-02,
  xmin=2018-07-02 13:00,
  xmax=2018-11-13 00:00,
  ymin=0,
  xtick={2018-07-02 13:00:00, 2018-07-11 13:00:00, 2018-08-02 18:00, 2018-08-17 12:00:00,2018-09-01 00:00:00, 2018-09-19 12:30:00,2018-10-01 00:00:00, 2018-10-23 12:40:00,2018-10-29 02:10:00, 2018-11-13 00:00},
xticklabel style={
                align=center,
                rotate=45,
                anchor=near xticklabel,
                yshift=0.0,
                below left,
                },
    xticklabels={\month\slash\day, Update 1.1.2 \\ (\month\slash\day), EoS 1.1.0  \\ (\month\slash\day), Update 2.0.0  \\ (\month\slash\day), \month\slash\day, EoS 1.1.1  \\ (\month\slash\day), \month\slash\day, EoS 1.1.2, Sapling  \\ (\month\slash\day),\month\slash\day},
    x label style={at={(axis description cs:0.5,-0.02)}},
    y label style={at={(axis description cs:0.03,0.5)}},
    xtick pos=left,
    ytick pos=left,
    legend style={draw=none},
    legend pos=outer north east,
    legend columns=1,
    stack plots=y,
    area style,
    enlarge x limits=false,
    width=0.94\textwidth,
    height=2.5in
 ]

 \addplot+[black, fill=sron0, each nth point=40, filter discard warning=false] table [col sep=comma,x=date time,y=1.1.0] {./sub2.csv}\closedcycle;
 \addplot+[black, fill=sron1, each nth point=40, filter discard warning=false] table [col sep=comma,x=date time,y=1.1.1] {./sub2.csv}\closedcycle;
 \addplot+[black, fill=sron2, each nth point=40, filter discard warning=false] table [col sep=comma,x=date time,y=1.1.2] {./sub2.csv}\closedcycle;
 \addplot+[sron3, fill=sron3, each nth point=40, filter discard warning=false] table [col sep=comma,x=date time,y=2.0.0] {./sub2.csv}\closedcycle;
 \addplot+[black, fill=sron4, each nth point=40, filter discard warning=false] table [col sep=comma,x=date time,y=2.0.1] {./sub2.csv}\closedcycle;
 \addplot+[black, fill=sron5, dashed, each nth point=40, filter discard warning=false] table [col sep=comma,x=date time,y=other] {./sub2.csv}\closedcycle;

 \legend{1.1.0,1.1.1,1.1.2,2.0.0,2.0.1,other}

\end{axis}
\end{tikzpicture}
\caption{Distribution of announced version strings in the Zcash peer-to-peer network.}
\vspace{-1.5em}
\label{fig:subversion}
\end{figure*}
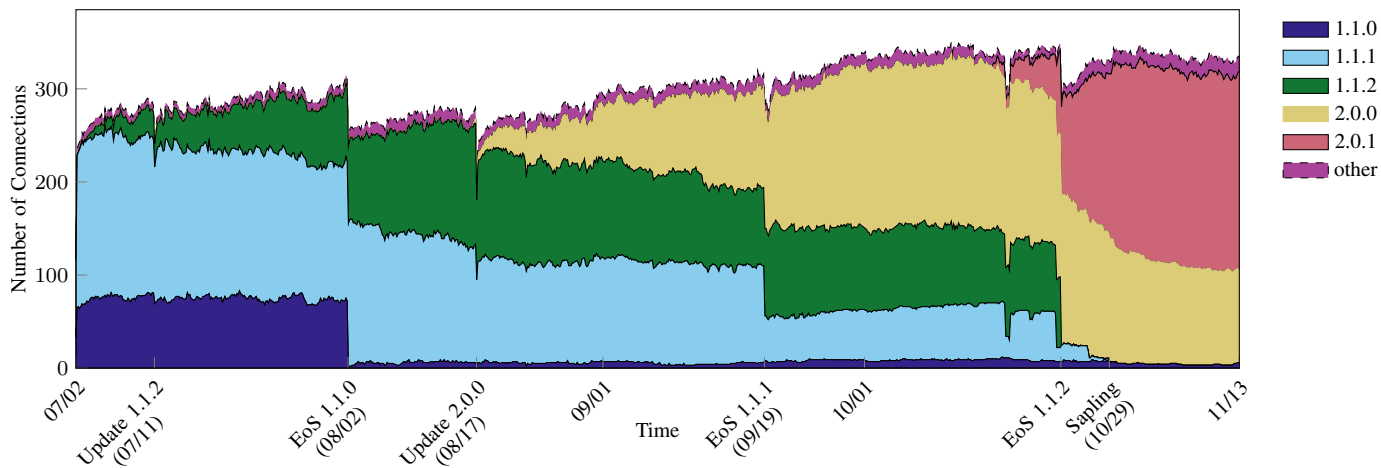

\subsection{Measurement Setup}
In order to characterize the Zcash network,
we deployed a Zcash reference client for data collection
as an observation point. The client was run on a virtual machine in our
university network (TU Berlin, IPv4).
However, minor client modifications were necessary to ensure a reliable
measurement setup that avoids unwanted side-effects and yields representative
results. For example, in order to get a comprehensive overview,
it is necessary that the observation point is connected to as many peers as
possible. We therefore increased the maximum number of outbound connections
and allowed for 873 simultaneous connections, which is the upper limit
provided by the reference client on a Linux machine.  Moreover, we disabled
the sending of block inventory messages that announce new blocks to peers.
We however recorded arrival times of block inventory messages from neighboring peers
to get an overview of the block propagation times in the network.
Additionally, we recorded information about connected peers and the mining difficulty.
In our measurements, we omitted all data points recorded during the initial blockchain download,
because this is a one-time phase and therefore does not represent normal operation.
The described methodology is on par with other measurement studies~\cite{neudecker2019}.

We recorded data snapshots about peers and blocks every five minutes and two
minutes, respectively.
While a time interval of two minutes does not guarantee to record every single block,
it should capture the vast majority of blocks\footnote{At the time of writing,
Zcash's block interval time is 2.5 minutes}.
In addition to the application layer data,
we sent ICMP ping messages to the connected peers,
which serve as a baseline and help to reveal networking problems during the measurements.

Our measurement period spans from July 2, 2018 until November 12, 2018.
It thus captures the time after Zcash's first network upgrade called \enquote{Overwinter}~\cite{zecNetwork}.
During the measurements, we upgraded the client twice to include new functionalities
and to ensure compatibility after Zcash's second network upgrade called \enquote{Sapling}~\cite{zecNetwork}.
Accordingly, client versions MagicBean~1.1.1, 1.1.2, and 2.0.0 were utilized.

\subsection{Measurement Results}

\paragraph*{Network size and client versions}
To begin with, we are interested in the network size and the number of nodes
running a specific software version.
In order to estimate these numbers, we count the number of network peers that
were connected to our measurement node over time.

The number of simultaneous connections held by our vantage point and
the distribution of client versions are shown in
\figurename~\ref{fig:subversion}. As the number of observed simultaneous
connections never exceeds this range, we estimate the size of the network to
be around 300 to 350 nodes. Note however that we observed 4,208 distinct IP
addresses which established a connection to our vantage point at least once.
In total, the address manager of our node learned from exchanged address information about
around 25,000 IP addresses. We assume that this high number of (stale) IP addresses
stems from a small number of network peers that constantly change their IP address.

As a side note, on November 11, we observed 117 simultaneous connections
coming from the same IP address.  These 117 connections were open for
approximately 30 minutes but did not send any data. The version name of this
client was \enquote{xbadprobe}, which suggests that this may have been an attempt to
occupy the node's inbound connections.

The reference client provided by the Zcash developer team uses the string \enquote{MagicBean} followed by a
version number as the client name. While it is possible to modify the client
software without changing the announced version string, we consider this
a negligible side-effect and use the strings to account for different
client versions. Likewise, the data shows that only a small fraction of clients use
custom version strings or minor versions. We therefore only
consider major client versions and categorize minor versions (\eg, 2.0.1-rc1)
and custom version strings as \enquote{other clients} in the following.

Since client version 1.0.9, each reference client comes with an End-of-Support
(EoS) period, which causes the client software to halt after the EoS period has expired.
At the time of writing, the period is set to be 64,512 blocks or approximately 16
weeks. Visually, we can clearly identify the EoS periods in
\figurename~\ref{fig:subversion}. For example, we captured the complete 16
week lifecycle for client version 1.1.2, showing the adoption of and transition
to other versions. Moreover, client versions older than 1.1.0
are no longer supported after the second network upgrade. Connections to
these outdated clients are therefore dropped after the version handshake.
Nevertheless, our data reveals that some clients were still active with older
client versions, even though they no longer supported the consensus.
However, the client versions 1.1.0 and 1.1.1 included a
configuration option which disabled the automatic halt after the EoS period,
which explains why the observed drop in numbers for those versions is not as
entire as for version 1.1.2.
The introduction of EoS periods is an interesting and clearly effective
approach to ensure that all users update
their client software and obsolete versions leave the network in a timely
manner. This reduces the threat that
bugs persist even after they were already fixed, which in turn bears the
danger to undermine the trust in the currency.

\paragraph*{Geographical and mining distribution} In \figurename~\ref{fig:countries},
we show the country distribution of observed
IP addresses and block origins during our measurement period, which provides insights
on Zcash's global adoption and mining power distribution. We used GeoIP~\cite{geoipdb} to map the IP addresses from our data set to
countries.  If a country had less than 30 IP addresses it is grouped as
\enquote{other}.  We further assume that the host which first announces the
block is also the block's miner.

We generally can see that the Zcash network spreads over all northern
continents.  From a country perspective, the clients are mainly present in
Russia and the United States.  From a continent perspective, however, the
network is evenly distributed over Europe, America, and Asia.  Only a small
number of nodes are located in the southern continents.

In contrast, the mining power distribution clearly shows a skewed
distribution, suggesting an immense centralization of mining power. We observe that
around 51\,\% of the blocks are created by 16 miners, out of which ten are located
in China. Overall, 53\,\% of all observed blocks originate from China, 10\,\% from France,
9\,\% from the United States, 5\,\% from Germany, and 4\,\% from the Netherlands.
The remaining 19\,\% are from different countries. Notably, while we observed only seven
IP addresses from Ireland, the country contributes 3\% of all blocks. It
should also be mentioned that we found some nodes that did not announce any
blocks.

We also observed a quadruplication in the difficulty during our measurements,
resulting from an increased mining power,
which effectively raises the bar to start mining blocks in the Zcash network.
This increase in mining power hence makes it unlikely that the country distribution
of miners will change in the near future. As new Equihash-capable ASIC mining hardware was
introduced in mid 2018, the increase in mining power is likely a result of a gradual change from GPU mining to ASIC mining.

In general, geographical centralization of mining power in one jurisdictional
area creates the risk of interference by a state actor. The
centralization of mining power in China we observed is in line with results from other
cryptocurrencies, \eg, Bitcoin~\cite{kaiser2018looming}.

\begin{figure}
\begin{tikzpicture}
  \footnotesize
  \begin{axis}[
  ymin=0,
  ybar=0pt,
  ylabel=Share,
  xlabel=Country,
  x label style={at={(axis description cs:0.5,-0.1)}},
  y label style={at={(axis description cs:0.05,0.5)}},
  xtick=data,
  xticklabels from table={\datatab}{country},
  xtick pos=left,
  ytick pos=left,
  x tick label style={rotate=90},
  bar width=3.0,
  enlarge x limits=0.03,
  no marks,
  height=4.5cm,
  width=\columnwidth,
  legend cell align={right},
      ]
  \addplot+[sron0,fill=sron0] table [x expr=\coordindex,y=ip,col sep=comma]{./countries_ip.csv};
  \addplot+[sron1,fill=sron1] table [x expr=\coordindex,y=block,col sep=comma]{./countries_ip.csv};
  \legend{Peers, Blocks}
  \end{axis}
  \end{tikzpicture}
  \vspace{-1ex}
  \caption{Distribution of observed addresses and block origins.}
  \label{fig:countries}
  \vspace{1em}

\begin{tikzpicture}
  \footnotesize
  \begin{axis}[
      xlabel=Delay in ms,
      ylabel=Share,
xmin=0,
      ymin=0,
      xmax=5000,
      ymax=0.013,
      minor grid style={gray!25},
      major grid style={gray!25},
      enlargelimits=false,
      xtick pos=left,
      ytick pos=left,
      height=4.5cm,
      width=\columnwidth,
      y label style={at={(axis description cs:0.05,0.5)}},
      ]
    \addplot+[sron0,ybar interval,mark=no] table[x=time,y=share,col sep=comma]{./block.csv};
\addplot+[gray,sharp plot, no marks,update limits=false] coordinates {(685,0) (685,1)}
     node[above,black] at (axis cs:940,0.011) {50~\%};
    \addplot+[gray,sharp plot, no marks,update limits=false] coordinates {(2000,0) (2000,1)}
     node[above,black] at (axis cs:2250,0.011) {90~\%};
  \end{axis}
  \end{tikzpicture}
  \vspace{-1ex}
  \caption{Block inventory arrival time differences.}
  \label{fig:blockpropagation}
  \vspace{1em}

\begin{tikzpicture}
  \footnotesize
  \begin{axis}[
    xlabel=Time,
    ylabel=Median Latency in ms,
    date coordinates in=x,
    date ZERO=2018-07-02,
    xmin=2018-07-02 00:00,
    xmax=2018-11-13 00:00,
    ymin=20,
    height=4cm,
    width=\columnwidth,
xticklabel style={
anchor=near xticklabel,
                    yshift=-5.0,},
    xticklabel=\month\slash\day,
    xtick pos=left,
    ytick pos=left,
    x label style={at={(axis description cs:0.5,-0.02)}},
    y label style={at={(axis description cs:0.05,0.5)}},
    legend pos=south west,
    legend style={legend columns=-1},
    ]

    \addplot+[sron0,only marks, mark size=0.5pt,each nth point=25, filter discard warning=false] table [col sep=comma,x=date time,y=ICMP-Median] {./latency.csv};
    \addplot+[sron1,only marks, mark size=0.5pt,each nth point=25, filter discard warning=false] table [col sep=comma,x=date time,y=ZEC-Median] {./latency.csv};
    \legend{ICMP, ZEC}
    \end{axis}
    \end{tikzpicture}
  \vspace{-1ex}
  \caption{Latencies of the connected clients.}
  \vspace{-1.5em}
  \label{fig:latencies}
\end{figure}

\paragraph*{Block propagation}
In order to determine the time it takes for a block to propagate in the network,
we measured the time it takes until a block is announced by all neighbors of our vantage point.
\figurename~\ref{fig:blockpropagation} accordingly shows the time differences
as a mass function of the first and all following observed block announcements (\ie, arrival of block inventory messages). In order to estimate when the neighbor peers learned about the blocks, the
shown times are adjusted by subtracting half RTT retrieved from Zcash's keep-alive messages.
The RTT is estimated using the exponential weighted moving average (EWMA) approach known from
TCP (cf.\ RFC 6298).

We note that block propagation follows a long-tailed distribution,
where a considerable number of inventory messages take significantly longer.
After 690\,ms, 50\,\% of all block inventory messages have arrived. And after
two seconds, 90\,\% of all nodes know the block. Furthermore, it takes a small number
of nodes a really long time to retrieve and propagate new blocks.
Interestingly, this is in accordance with similar observations ($\approx2\,s$
for 90\,\% coverage) made in~\cite{neudecker2019} for the Bitcoin network,
even though the latter is considerably larger (at the time of this paper: $\approx350$ vs. $\approx10.000$ nodes).

In general, we can say the network needs around 700~ms for a block to be known by most peers
and roughly two seconds to spread the information in the whole network.
However, this delay does not seem to have a significant negative impact on
consensus.
From the 57,365 observed block hashes, only 297 are not included in the Zcash blockchain,
yielding a stale block rate of 0.337\,\%. This measurement is comparable to
the stale rates found in other cryptocurrency networks like Bitcoin or
Litecoin, which according to~\cite{gervais2016security} exhibited around
0.41\,\% and 0.273\,\% stale blocks, respectively.

\paragraph*{Network stability}
Lastly, we are interested in the network load and stability.
\figurename~\ref{fig:latencies} shows the median latencies to all connected nodes over time.
As a baseline, we additionally measured ICMP ping times and compare them to
the Zcash ping measurements. The Zcash ping is a keep-alive message, scheduled every two minutes,
which is send to all neighbors via the respective TCP connection
and is usually processed with all other exchanged messages. This leads to
head-of-line blocking during transaction and block relay, whereby the Zcash
ping messages are delayed. In comparison to the more reliable ICMP ping, it
therefore rather serves as an indicator of a node's activity.

The ICMP ping's median values vary between 45\,ms and 55\,ms
and are lower and less fluctuating than Zcash pings.
From October 18, the median ICMP ping and Zcash ping latencies increased.
This behavior could be a reaction to the second upgrade (\enquote{Sapling}).
In this upgrade, the performance of the so-called shielded transactions was
improved, which might have made them more attractive and could have lead to increased usage.
The increase of the ICMP ping could also be due to network interferences in the University network.

As in any other open peer-to-peer network, nodes in the Zcash network can join and leave at any time,
which results in an ever-changing topology.
In order to assess the network stability, we analyzed the lifetime of
connections. In order to circumvent artificial spikes in our data set stemming from our own client
updates, we only consider the time after we updated our measurement node to
client version 2.0.0.

We generally observed that around 50\,\% of connections remained active for at
least 50 minutes. Moreover, around 20\,\% of connections lasted longer than a
day, while 10\,\% were active for more than four days, and 1\,\% have even
remained active since the beginning of our study. We also observed around 24\,\%
of short-lived connections, which were active less than five minutes.
In summary, we could see that a good part of the network consists of long-lived
stable peers. This \emph{core network} can certainly become a reasonable target for topology inference.

 \section{Topology Inference}\label{sec:inference}
In the following, we aim to infer the interconnectivity of Zcash network peers
by conducting a passive timing analysis, which allows us to incrementally uncover the topology
of the peer-to-peer network.
To this end, we build upon and extend the inference model developed by
\citeauthor{neudecker2016timing}~\cite{neudecker2016timing,neudecker2019network}.
We however replace the employed active transaction measurements with passive block measurements,
which, as we will see, has a number of advantages.
In a nutshell, the goal is to infer the connection between two peers
by monitoring when they emit inventory messages announcing new blocks.
The model concentrates on detecting connections between peers adjacent to
mining nodes, which can be used to reveal security-critical information of the network.

\subsection{Inference Model}
Let us assume a source node $S$, which we consider the origin of a block,
and a relay node $R$.
Moreover, assume that node $S$ and $R$ are both interconnected
and also connected to a measurement node $M$.

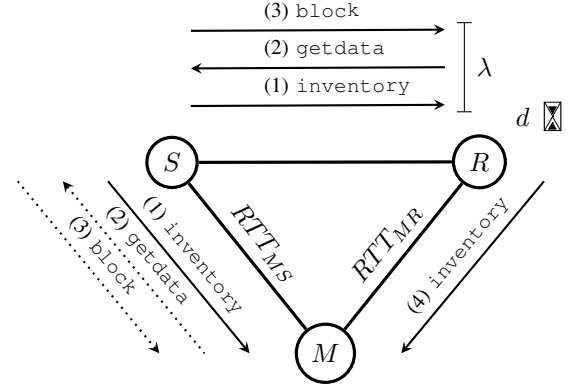
\begin{figure}[!t]
  \centering

  \begin{tikzpicture}[node distance=20mm and 15mm, >=stealth]
\tikzstyle{node}=[circle,draw,very thick]
    \tikzstyle{label}=[font=\footnotesize]
    \node[node] (s) {$S$};
    \node[node,below right=of s] (m) {$M$};
    \node[node,above right=of m] (r) {$R$};

    \draw[very thick] (s) -- (r);
    \draw[very thick] (s) -- (m) node [above,midway,sloped] {$RTT_{MS}$};
    \draw[very thick] (r) -- (m) node [above,midway,sloped] {$RTT_{MR}$};

\coordinate[xshift=-6mm] (s-left1) at (s.south west);
    \coordinate[xshift=-6mm] (m-left1) at (m.west);
    \draw[->, thick] (s-left1) -- (m-left1)
      node [above,midway,sloped,label] {(1) \texttt{inventory}};

\coordinate[xshift=-6mm] (s-left2) at (s-left1);
    \coordinate[xshift=-6mm] (m-left2) at (m-left1);
    \draw[<-,dotted,thick] (s-left2) -- (m-left2) node [above,midway,sloped,label] {(2) \texttt{getdata}};

\coordinate[xshift=-6mm] (s-left3) at (s-left2);
    \coordinate[xshift=-6mm] (m-left3) at (m-left2);
    \draw[->,dotted,thick] (s-left3) -- (m-left3) node [above,midway,sloped,label] {(3) \texttt{block}};

\coordinate[xshift=6mm] (r-right1) at (r.south east);
    \coordinate[xshift=6mm] (m-right1) at (m.east);
    \draw[->,thick] (r-right1) -- (m-right1)
      node [above,midway,sloped,label] {(4) \texttt{inventory}};

\coordinate[yshift=5mm] (s-top1) at (s.north east);
    \coordinate[yshift=5mm,xshift=-2mm] (r-top1) at (r.north west);
    \draw[->,thick] (s-top1) -- (r-top1) node [above,near start,anchor=south west,sloped,label] {(1) \texttt{inventory}};

\coordinate[yshift=5mm] (s-top2) at (s-top1);
    \coordinate[yshift=5mm] (r-top2) at (r-top1);
    \draw[<-,thick] (s-top2) -- (r-top2) node [above,near start,anchor=south west,sloped,label] {(2) \texttt{getdata}};

\coordinate[yshift=5mm] (s-top3) at (s-top2);
    \coordinate[yshift=5mm] (r-top3) at (r-top2);
    \draw[->,thick] (s-top3) -- (r-top3) node [above,near start,anchor=south west,sloped,label] {(3) \texttt{block}};

\coordinate[xshift=-2mm] (r0) at (r);
    \draw[|-|,shorten >=-1mm,shorten <=-1mm] (r0 |- r-top1) -- (r0 |- r-top3);
    \node[anchor=west,xshift=3mm] at (r-top2) {$\lambda$};

\node[above right=1mm and 1mm of r] (d) {$d$};
    \node[right=-2pt of d] {\includegraphics[height=\baselineskip]{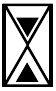}};
  \end{tikzpicture}

  \caption{Block propagation with three nodes.}
  \vspace{-1em}
  \label{fig:blck}
\end{figure}

A schematic of this three-node scenario is shown in \figurename~\ref{fig:blck}.
While numbered arrows indicate the order and direction in which messages are
sent, dashed arrows indicate further protocol messages that are
however only shown for completeness, but do not influence the inference model.
Hence, a block relay in Zcash consists of a three-way message exchange
(announce, request, response):
An \texttt{inventory} message announces a block,
a \texttt{getdata} message requests the block,
and a \texttt{block} message transmits the actual block data.

From the perspective of $M$, we have knowledge about the time node $S$ and $R$
announced a specific block to node $M$.
We also know the round-trip times $RTT_{MS}$ and $RTT_{MR}$
between nodes $M$ and $S$, as well as nodes $M$ and $R$, respectively.
The measured arrival times consist of at least a delay introduced by
the latency between $S$ and $R$ and the processing delay of $R$.
Since the measurements are conducted on $M$, the measured time
includes $RTT_{MS}/2$ as well as $RTT_{MR}/2$, which we subtract once
prior to the following calculations.
Note that if the latency difference is \enquote{small enough},
$S$ and $R$ are directly connected with a high probability.
We take this as the basis for our timing analysis.
In the following, we will derive our model in detail.

When we assume that the link latency between any two peers follows the same distribution $\lambda$
and a node's processing delay can be described as $d$,
then the probability of a time difference $t$
with $h$ edges in between the two reference nodes is given by
\begin{equation}\label{eq:pth}
P(\Delta=t|H=h)= (\lambda^{*h} * d^{*h})(t).
\end{equation}
Please note that the $*$-operator denotes a convolution;
accordingly, $\lambda^{*h}$ and $d^{*h}$ denote the $h$ convolution power.
To infer the topology, we have to calculate the probability of $h$ edges
assuming a time difference $t$.
This is possible using Bayes' Theorem,
\begin{equation}\label{eq:pht}
P(H=h|\Delta=t)=\frac{P(\Delta=t|H=h) \cdot P(H=h)}{P(\Delta=t)}.
\end{equation}

The probability that an inventory message arrives after time~$t$, $P(\Delta=t)$,
can be calculated according to the law of total probability.
The probability $P(\Delta=t|H=h)$ is given by~\eqref{eq:pth}.
The probability of $h$ edges between the two reference nodes, $P(H=h)$,
can be calculated by assuming an Erd\H{o}s-R\'enyi random graph model,
where the probability of an edge is calculated based on the mean degree.
In the case of a mean degree $deg$ and $N$ nodes, the equation is given by
\begin{equation}\label{eq:ph}
P(H=h)=\biggl[1-\biggl( \frac{deg}{N-1} \biggr)\biggr]^{h-1} \cdot \biggl( \frac{deg}{N-1} \biggr).
\end{equation}

For our model,
we assume that the link latency distribution $\lambda$ is a result of the
three-way message exchange to relay a block.
Furthermore, we assume that this latency distribution follows a normal distribution,
\ie, $\lambda(x)=\mathcal{N}(x;\mu_{\lambda},\sigma_{\lambda}^{2})$.
The expected propagation time is presumed to depend on the geolocation of $S$ and $R$.
While this is a very simplifying assumption as latencies may depend on many factors,
including AS and peering relationships,
country-based measurements are already available and/or easier to obtain in decent quality.
The mean~$\mu_{\lambda}$ and variance $\sigma_{\lambda}^{2}$ values of the normal distribution
can accordingly be estimated by using RTT measurements for the respective
geolocations, multiplied by a factor of $1.5$ to mind three-way message exchange.

We define a node's processing delay $d$ as the sum of
transmission delay, queuing delay, and block verification time.
While the transmission delay can have a significant impact for very large block sizes,
we assume the block verification time to be the dominating factor. We
therefore assume the network-based delays to be adequately captured by the link latency and
leave the development of an advanced transmission model as an
open question for future research.
Furthermore, we generally assume a linear correlation
between the time it takes to validate a new block and the number of included
transactions, \ie, the block size $s_b$.
However, since the network peers run on equipment of varying power, the
appropriate validation factor is not necessarily a global constant. To account for
variations, we model the processing delay $d$ as a normal distribution,
\ie, $d(x)=\mathcal{N}(x;\mu_{d},\sigma_{d}^{2})$.
The resulting function is
\begin{equation}
P(t-\epsilon \leq t \leq t+\epsilon |H=h)= \int_{t-\epsilon}^{t+\epsilon}\mathcal{N}\bigl(t;\mu,\sigma^{2}\bigr)dt
\end{equation}
with $\mu=h\cdot(\mu_{\lambda}+\mu_{d})$,
$\sigma^{2}=h\cdot(\sigma_{\lambda}^{2}+\sigma_{d}^{2})$.
Here $\epsilon$ is a tolerance variable adjusting for the possibility of measurement errors.
Given the complexity of this model, the selection of reasonable mean and
variance values is important to reach an adequate degree of accuracy.

\subsection{Parametrization}
The inference model requires a number of parameters which have an impact on the
accuracy of the resulting estimations.
These parameters include the normal distributions for the latency~$\lambda$
and the processing delay~$d$, as well as the value for the tolerance variable~$\epsilon$.

For the latency distribution~$\lambda$, different data sources are possible.
We consider the iPlane dataset~\cite{iplane},
which provides publicly available data of global latency measurements,
as a viable data source.
Admittedly, the data is somewhat outdated but still fits the purpose.
It consist of pairs of globally distributed IP addresses and a
corresponding RTT value measured on a specific point in time.
As an alternative data source, we suggest to conduct ICMP ping measurements
from the observation points
or to directly utilize the Zcash ping measurements (cf. \figurename~\ref{fig:latencies}).

Values for the validation time can also be acquired through different means.
We consider the evaluation constant from~\cite{gervais2016security}
as a reasonable estimation for~$\mu_{d}$.
Unfortunately, in this case we must make some assumptions on the variance.
Alternatively, $\mu_{d}$ and $\sigma_{d}^2$ can be acquired experimentally,
\ie, by averaging the processing time of a local Zcash node.
We compare both approaches in our evaluation.

Varying the tolerance parameter~$\epsilon$ should generally have
no significant influence on the measurement results.
However, larger values for $\epsilon$ increase the influence of the likelihood
$P(t-\epsilon \leq t \leq t+\epsilon |H=h)$
and therefore decrease the influence of the prior probability $P(H=h)$
on the posterior probability.
 \section{Evaluation}\label{sec:evaluation}
We evaluate our inference model using two different ways:
through a simulation scenario and a real world measurement test with two Zcash nodes.

\subsection{Methodology}
In order to facilitate the simulation-based evaluation scenario, we created an
undirected graph with a certain amount of vertices utilizing the Boost Graph Library~\cite{bgl}.
Every vertex represents a Zcash networking node and creates 8
edges to randomly selected vertices, representing the 8 outgoing
connections the reference client tries to establish. The mean degree of the
random topology is therefore 16. We assign a country to each vertex
and draw an according link latency value for each edge from a normal distribution
that was parametrized with the iPlane data set~\cite{iplane}. As
discussed before, we multiply this latency value by factor 1.5 to mind the
entire three-way message exchange. Moreover, we simulate our vantage point which
is connected to all other vertices and therefore able to observe the
simulated propagation behavior in its entirety.

Given this network graph, we can deduce simulated time difference measurements
by traversing the shortest path between any node and our vantage
point, accumulating the edge weights accordingly.
In order to retrieve statistically significant results, we repeated this process
50 times for varying edge weights and applied our topology inference model for each of
the measurements. Each time, we calculated the probability under a tolerance
of $\epsilon=5\,ms$ for possible hop counts $h$ of 1 to 9 and
finally calculated the mean value for each distance. From this, we deem the
value with highest mean probability as the likeliest estimated
distance. As we are in full knowledge of the simulated topology, this scenario
allows us to exactly determine the precision and recall of our model under the
influence of different parametrizations for the processing time distribution
$d = \mathcal{N}(x;\mu_d,\sigma_d^2)$ and block sizes $s_b$. The validation
constants $k_\mu$ and $k_{\sigma^2}$ determine the distribution parameters as
$\mu_d = k_\mu \cdot s_b$ and $\sigma_d^2 = k_{\sigma^2} \cdot s_b$.

Theoretically, our model could predict the probability of $h$ up to
the network diameter. However, our model becomes less accurate for
distances above three, since path lengths larger than two exhibit an increased
possibility for parallel running paths. Therefore, our model is suited best to
infer individual connections one by one.
Additionally, as our model utilizes assumptions about the geographic locations
of network peers to determine their edge latencies, inaccuracies in the geographic clustering
can result in high amount of false positives for certain peer connections. For
example, if we assume a scenario in which two nodes located in Germany are connected via a
third node located in Russia, our model could
predict a distance of five (low latency) edges, even though the simulated path
consists only of three (high latency) links.

In our simulation, we considered topologies with 300 nodes, which resembles
the measured network size. Furthermore, we evaluate our model in the
simulation for distances up to three. For the geographic
distribution, we chose two countries each from America and Europe, and three
countries from Asia, whereby we aim to represent the northern Hemisphere. The
distribution assigned as follows: 30\,\% United States, 20\,\% in Russia,
10\,\% Canada, 10\,\% China, 10\,\% France, 10\,\% Germany, and 10\,\% Japan.

\subsection{Simulation Results}
We use a simulation scenario to evaluate different parametrizations of
the processing time distribution $d$. That
is, we evaluated precision and recall of our model for different block sizes $s_b$
assuming the validation constants $k_\mu=0.3796\,\mu s\slash B$ and
$k_{\sigma^2}=0.552049\,\mu s^2\slash B$ taken from~\cite{gervais2016security}.
We also ran the simulations for three different degrees of latency standard deviation: \enquote{small},
\enquote{medium} and \enquote{large}, assuming 10\,\%, 30\,\%, and 50\,\% of
the mean, respectively.

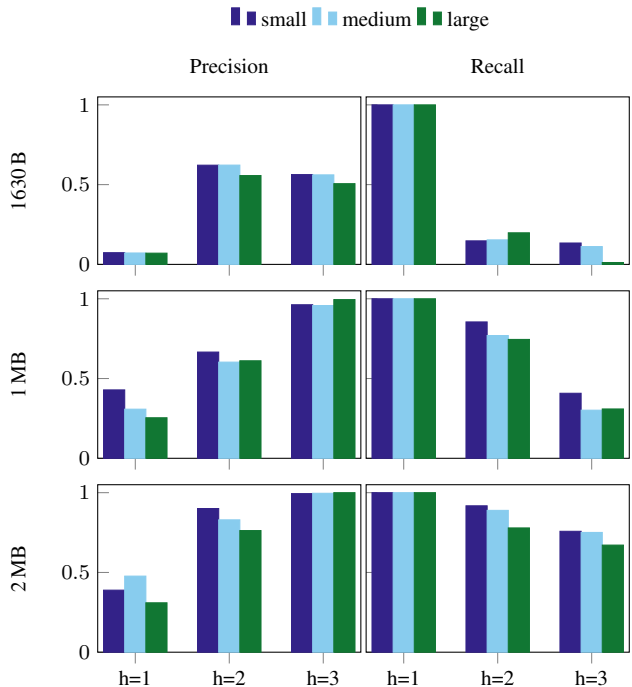
\begin{figure}
 \begin{tikzpicture}
    \footnotesize
    \begin{axis}[
    ymin=0,
    ymax=1,
    ybar=0pt,
    enlarge x limits=0.2,
    no marks,
    hide axis,
    xmin=0,
    xmax=50,
    legend columns=3,
    width=3.45in,
    height=1.9in,
    legend style={at={(0.75,1)},anchor=east, draw=none}
    ]
    \addlegendimage{sron0, fill=sron0,}\addlegendentry{small};
    \addlegendimage{sron1, fill=sron1,}\addlegendentry{medium};
    \addlegendimage{sron2, fill=sron2,}\addlegendentry{large};
    \end{axis}

    \begin{groupplot} [
    group style={group name=spr,group size=2 by 4, x descriptions at=edge bottom, y descriptions at=edge left, horizontal sep=2pt, vertical sep=10pt,},
    ymin=0,
    ymax=1.05,
    width=2in,
    height=1.5in,
    ybar=0pt,
    enlarge x limits=0.2,
    no marks,
    xtick=data,
    xtick pos=left,
    ytick pos=left,
    xticklabels from table={\datatabelle}{hop},
    y label style={at={(axis description cs:0.05,0.5)}},
    ]
    \nextgroupplot[
        bar width=8.0,
        title=Precision,ylabel=1630\,B]
        \addplot+[sron0,fill=sron0] table [x expr=\coordindex,y=p,y error=perr,col sep=comma]{./small_1630_country.csv};
        \addplot+[sron1,fill=sron1] table [x expr=\coordindex,y=p,y error=perr, col sep=comma]{./medium_1630_country.csv};
        \addplot+[sron2,fill=sron2] table [x expr=\coordindex,y=p,y error=perr, col sep=comma]{./big_1630_country.csv};
    \nextgroupplot[bar width=8.0,title=Recall,]
        \addplot+[sron0,fill=sron0] table [x expr=\coordindex,y=r,y error=rerr,col sep=comma]{./small_1630_country.csv};
        \addplot+[sron1,fill=sron1] table [x expr=\coordindex,y=r,y error=rerr, col sep=comma]{./medium_1630_country.csv};
        \addplot+[sron2,fill=sron2] table [x expr=\coordindex,y=r,y error=rerr, col sep=comma]{./big_1630_country.csv};
    \nextgroupplot[bar width=8.0,ylabel=1\,MB]
        \addplot+[sron0,fill=sron0] table [x expr=\coordindex,y=p,y error=perr,col sep=comma]{./small_1MB_country.csv};
        \addplot+[sron1,fill=sron1] table [x expr=\coordindex,y=p,y error=perr, col sep=comma]{./medium_1MB_country.csv};
        \addplot+[sron2,fill=sron2] table [x expr=\coordindex,y=p,y error=perr, col sep=comma]{./big_1MB_country.csv};
    \nextgroupplot[bar width=8.0,]
        \addplot+[sron0,fill=sron0] table [x expr=\coordindex,y=r,y error=rerr,col sep=comma]{./small_1MB_country.csv};
        \addplot+[sron1,fill=sron1] table [x expr=\coordindex,y=r,y error=rerr, col sep=comma]{./medium_1MB_country.csv};
        \addplot+[sron2,fill=sron2] table [x expr=\coordindex,y=r,y error=rerr, col sep=comma]{./big_1MB_country.csv};
    \nextgroupplot[bar width=8.0, xtick=data, xticklabels from
    table={\datatabelle}{hop}, ylabel=2\,MB]
        \addplot+[sron0,fill=sron0] table [x expr=\coordindex,y=p,y error=perr,col sep=comma]{./small_2MB_country.csv};
        \addplot+[sron1,fill=sron1] table [x expr=\coordindex,y=p,y error=perr, col sep=comma]{./medium_2MB_country.csv};
        \addplot+[sron2,fill=sron2] table [x expr=\coordindex,y=p,y error=perr, col sep=comma]{./big_2MB_country.csv};
    \nextgroupplot[bar width=8.0, xtick=data, xticklabels from table={\datatabelle}{hop},]
        \addplot+[sron0,fill=sron0] table [x expr=\coordindex,y=r,y error=rerr,col sep=comma]{./small_2MB_country.csv};
        \addplot+[sron1,fill=sron1] table [x expr=\coordindex,y=r,y error=rerr, col sep=comma]{./medium_2MB_country.csv};
        \addplot+[sron2,fill=sron2] table [x expr=\coordindex,y=r,y error=rerr, col sep=comma]{./big_2MB_country.csv};
    \end{groupplot}
 \end{tikzpicture}
 \caption{Precision and Recall for different block sizes, different variances and 300 nodes located in 7 different countries.}
 \vspace{-1.5em}
  \label{fig:spr}
\end{figure}

The simulation results presented in \figurename~\ref{fig:spr} show that the
model generally performs better with larger block sizes, which yield a precision of up
to 40\,\% and recall of up to 100\,\%.
Surprisingly, the precision for a direct connection is slightly higher for a 1\,MB block than for a
2\,MB block. However, the validation time seems to be a good indicator for the
number of hops, since the precision for distances two and three are above 50\,\%
Moreover, the larger the block sizes, the higher are the estimated processing
times. Hence, the lower is the share of the estimated latency, which makes the
model less dependent on accurate latency estimations.

\subsection{Real-World Experiments}
In the real-world evaluation scenario, we deployed two nodes in the
peer-to-peer network over the course of one week.
One node functioned as the measurement node and recorded the arrival time of all block
inventory messages. Moreover, as a point of reference, the relay node
recorded its connections over the time of measurement.

Afterwards, we apply our model to infer if
a direct connection between the different block creators and the relay node
existed in the given time frame. We calculated RTT estimations
using the EWMA approach.

In the measurement period, we estimate the network size to be 316 nodes of
which 87 nodes sent at least five blocks. Overall, we recorded 4,160 blocks with an average block size of $15,678 \, B$ and 21 stale blocks. We also recorded the average
validation times our client exhibited during the verification of 5,000 blocks
from the Zcash main network. Resulting from this, we set the estimated validation constants to
$k_\mu=12.7357 \, \mu s\slash B$ and $k_{\sigma^{2}}=2128.16 \, \mu s^2\slash B$. 

For the further evaluation, we only consider direct connections with the
(mining) nodes that announced at least five blocks.

As shown in \tablename~\ref{tab:eval}, even despite the low block sizes, our
model is able to achieve a precision of 50\,\% and a recall of 82.5\,\% under
the discussed parametrization.

\begin{table}
\centering
\caption{Real-world measurement results.}
\begin{tabularx}{\columnwidth}{Xrrr}
\toprule
 & \textbf{Model of~\cite{gervais2016security}} & \textbf{Test net.} &
 \textbf{Main net.} \\
$k_{\mu}$ & $0.38 \, \mu s\slash B$ & $8.55 \, \mu s\slash B$ & $12.74 \, \mu s\slash B$ \\
$k_{\sigma^2}$ & $0.55 \, \mu s^2\slash B$ & $345.1 \, \mu s^2\slash B$
& $2128.16 \, \mu s^2\slash B$ \\
\midrule
True Positive & 18 & 30 & 33 \\
False Positive & 18 & 26 & 33 \\
False Negative & 22 & 10 & 7 \\
Precision & 50 \% & 53.5 \% & 50 \% \\
Recall & 45 \% & 75 \% & 82.5 \% \\
\bottomrule
\end{tabularx}
\vspace{-1.5em}
\label{tab:eval}
\end{table}

\subsection{Discussion}
This real-world evaluation scenario shows that we can consider half of the
inferred connections as correct and only a small amount of direct connections
as missing.
The results support the general validity of our inference model.
Considering the deliberate simplifying assumptions,
the approach seems to be a promising step
for inferring the topology of many cryptocurrencies similar to Bitcoin.
Furthermore, countermeasures for our method---\eg, artificially delaying block relays---would
increase the probability of stale blocks,
which weakens the system.
It therefore becomes clear that our method will remain usable in the future.

However, it should be possible to improve the precision of the current model by
acquiring more diversified latency measurements and considering additional
information about the client location, like the AS number. For higher block sizes, a
better modeling of the validation time and the transmission delay could also
improve the results.

Furthermore, the calculation of the probability of a certain amount of edges
between nodes ($P(H=h)$) is calculated based on an Erd\H{o}s-R\'enyi model,
which is a random topology. However, it is unlikely that nodes are connected
randomly in a peer-to-peer network in which certain nodes have a higher uptime
than others. It is more likely that some central nodes have more than 16
connections, especially nodes of the core network.
We leave these ideas open for future research.
 \section{Conclusion}\label{sec:conclusion}
In this work, we presented the first longitudinal measurement study of the Zcash
peer-to-peer network, revealing key characteristics of the
Zcash network.
Moreover, we introduced a topology inference model that relies on a passive
timing analysis of measured block arrival times. Our evaluation shows that this model allows us to infer
node interconnections with a precision of 50\,\% and recall of 82\,\% in the
real-world scenario. We deem this an attractive approach for passive topology inference of cryptocurrency networks.

\printbibliography
\end{document}